\documentclass[sigconf]{acmart}
\renewcommand\footnotetextcopyrightpermission[1]{}
\usepackage{float}
\AtBeginDocument{%
  }

\begin{document}

\title{DCS Chain: A Flexible Private Blockchain System}

\author{Jianwu Zheng}
\affiliation{%
  \institution{Shanghai Jiao Tong University}
  \city{Shanghai}
  \country{China}}
\email{jianwuzheng@sjtu.edu.cn}

\author{Siyuan Zhao}
\affiliation{%
  \institution{Shanghai Jiao Tong University}
  \city{Shanghai}
  \country{China}}
\email{zsy123@sjtu.edu.cn}

\author{Zheng Wang}
\affiliation{%
  \institution{Shanghai Jiao Tong University}
  \city{Shanghai}
  \country{China}
}
\email{wzheng@sjtu.edu.cn}

\author{Li Pan}
\affiliation{%
  \institution{Shanghai Jiao Tong University}
  \city{Shanghai}
  \country{China}}
\email{panli@sjtu.edu.cn}

\author{Jianhua Li}
\affiliation{%
  \institution{Shanghai Jiao Tong University}
  \city{Shanghai}
  \country{China}}
\email{lijh888@sjtu.edu.cn}

\begin{abstract}
  Blockchain technology has seen tremendous development over the past few years. Despite the emergence of numerous blockchain systems, they all suffer from various  limitations, which can all be attributed to the fundamental issue posed by the DCS trilemma. In light of this, this work introduces a novel private blockchain system named \textbf{\textit{DCS Chain}}. The core idea is to quantify the DCS metrics and dynamically adjust the blockchain's performance across these three dimensions, to achieve theoretically optimal system performance. Overall, our system provides a comprehensive suite of blockchain essentials, including DCS quantification, consensus protocol adjustment, and communication network simulation. 
  Our code is available at: https://github.com/zhengwang100/DCSChain.
\end{abstract}




\maketitle

\section{Introduction}
Since the inception of Bitcoin~\cite{nakamoto2008bitcoin}, blockchain technology has gained widespread attention in global research and practical applications. This pioneering distributed ledger technology not only forms the basis of digital currencies but also provides decentralized, immutable, transparent solutions for various fields. For example, in finance, blockchain has catalyzed the emergence of decentralized financial systems like decentralized exchanges~\cite{malamud2017decentralized} and decentralized finance~\cite{harvey2021defi}, empowering users to trade and manage assets independently of traditional financial institutions. As the technology evolves and its applications diversify, blockchain is increasingly recognized as a pivotal force for social and economic progress. Ongoing academic research consistently explores the technology, emphasizing its potential to play a significant role in future societies.

\begin{figure}[!t]
  \centering
  \includegraphics[width=\linewidth]{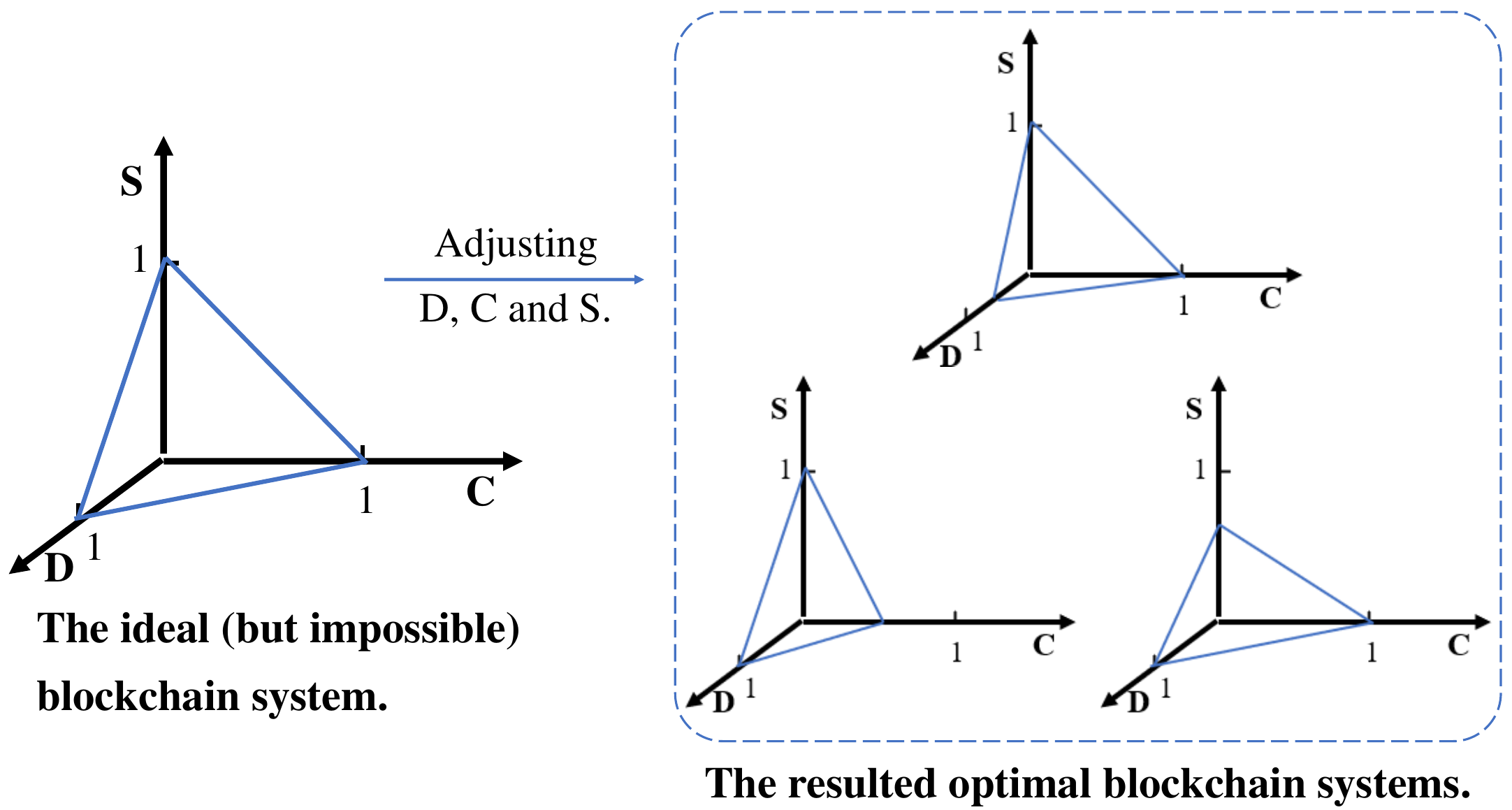}
  \caption{Based on the DCS trilemma, D, C and S should be dynamically adjusted to obtain the optimal blockchain performance.}
  \label{dcs_triangle}
\end{figure}

Despite unprecedented innovation and application potential brought by blockchain technology, existing blockchain systems generally reveal several limitations, such as insufficient throughput and inadequate decentralization~\cite{zheng2018blockchain,zeng2019review}. These limitations are primarily encompassed in what is known as the DCS trilemma/theory~\cite{zhang2018towards}. This trilemma suggests that it is impossible for a distributed  system to simultaneously achieve the three essential properties: Decentralization, Consistency, and Scalability. For instance, Bitcoin~\cite{nakamoto2008bitcoin} sacrifices scalability to achieve better decentralization while maintaining consistency. On the other hand, Hyperledger~\cite{androulaki2018hyperledger}, as a private blockchain, sacrifices decentralization to achieve better scalability while maintaining consistency.\footnote{Generally, blockchain systems prioritize consistency to a high degree, as compromising on consistency can lead to forks, which are generally considered unacceptable in blockchain technology.}

This paper introduces a flexible private blockchain system based on the DCS trilemma—\textbf{\textit{DCS Chain}}. As shown in Figure~\ref{dcs_triangle}, the core idea is to quantify and dynamically adjust the three DCS attributes to achieve the optimal blockchain system performance. Additionally, our system integrates some foundational modules such as local network simulation, key pre-distribution, and public key infrastructure to ensure secure and stable operations. Consequently, this system meets general private blockchain business needs and environmental adaptability, while offering new insights for the further development and innovative application of blockchain technology in private environments.

\section{System Overview}
\begin{figure}[t]
  \centering
  \includegraphics[width=\linewidth]{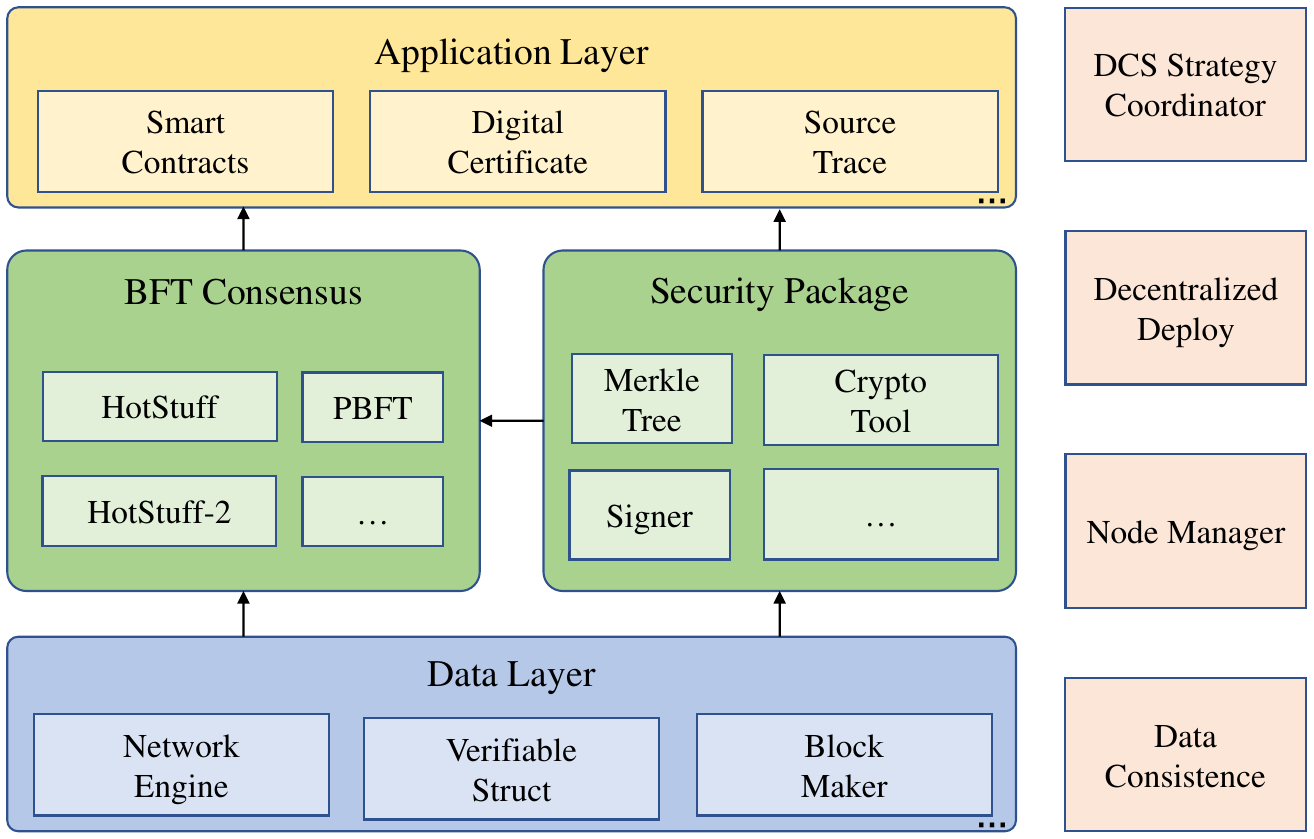}
  \caption{The architecture of DCS Chain}
  \label{arch}
  \Description{A woman and a girl in white dresses sit in an open car.}
\end{figure}
As illustrated in Figure \ref{arch}, our system offers a comprehensive suite of foundational components for private blockchains, including network simulation, consensus protocols, and cryptographic security suites. The core features are flowing: 1) the definition and quantification of the DCS trilemma indicators, and 2) a dynamic blockchain system adjustment mechanism based on these DCS indicators.
\subsection{Definition and Quantification of DCS}
The DCS trilemma~\cite{zhang2018towards} is pivotal in the conceptualization and refinement of blockchain systems. To this end, in the following, we define and quantify the DCS metrics, facilitating the dynamic adjustment of these metrics in subsequent system designs to enhance system performance.

\subsubsection{Definition of DCS}
We start by explaining the definitions of the three fundamental attributes in the DCS theory:
\begin{itemize}
\item \textbf{Decentralization}: Decentralization refers to the autonomy of a network from reliance on a trusted third party for its reliability. This attribute can be quantified by assessing the total number of nodes participating in decision-making processes and the degree of equity in their involvement.

\item \textbf{Consistency}: Consistency means that the blockchain data recorded by any node should be entirely uniform, with transactions, once committed, being immutable. The current state of the blockchain should be fully verifiable through the entirety of its historical records.

\item \textbf{Scalability}: Scalability is the ability of a blockchain system to increase its performance, including metrics such as throughput and low latency, proportionately with the augmentation of allocated resources, thereby enhancing both its capacity and usability.
\end{itemize}
\subsubsection{Quantification of DCS}
First, for the characteristic of decentralization, the data from consensus nodes are generally considered to be a critical consideration, implying that a higher number of consensus nodes correlates with a stronger degree of decentralization~\cite{9438771,gochhayat2020measuring}. Second, regarding consistency, there has been extensive research in the field of databases, which tends to quantify consistency as being inversely proportional to consensus latency~\cite{burckhardt2014principles,thomas1979majority}. Third, for scalability, throughput is always regarded as an important metric for systems related to data processing tasks, with the notion that the greater the throughput, the better the system's scalability~\cite{862209,chauhan2018blockchain}.

Based on the aforementioned discussion, we quantify $D_{rate}$, $C_{rate}$, and $S_{rate}$\text{------}the metrics of the three dimensions of DCS, respectively:
\begin{align}
    D_{rate} & = 1-\frac{1}{n} \\
    C_{rate} & = \frac{1}{e^{t}} \\
    S_{rate} & = 1-\frac{1}{ \lg{\theta} }
\end{align}
where $n$ represents the total number of consensus nodes within the blockchain network, $t$ denotes the latency experienced by a transaction from initiation to completion, and $\theta$ signifies the throughput of the blockchain system. 
Here, for practical purposes, we have also normalized the DCS metrics within the interval of 0 to 1 by introducing appropriate proportional and inverse proportional functions.

\subsection{Dynamic Adjustment Mechanism}
The dynamic adjustment mechanism is the core feature for achieving optimal performance under the guidance of the DCS theory. To this end, we have designed a flexible, multidimensional adjustment strategy that optimizes the number of consensus nodes, the type of consensus protocol, and the batch size, thereby achieving theoretically optimal blockchain performance.

\subsubsection{Number of Consensus Nodes}
The number of nodes participating in Byzantine fault-tolerant (BFT) consensus significantly impacts blockchain performance. As the number of nodes increases, system fault tolerance improves. However, each node must communicate with others to reach consistency, resulting in increased communication overhead and computational resource demands, which can reduce system throughput. Additionally, the time required for nodes to receive and process messages from others increases, leading to higher latency for each consensus round. For example, in the Practical Byzantine Fault Tolerance (PBFT~\cite{pbft}) algorithm, the maximum tolerable number of Byzantine nodes is $f$. If the number of nodes increases from $n$ to $2n$, each node has to wait for $2n-2f$ messages during the commit phase, compared to $n-f$ messages previously. This increase results in reduced throughput and increased latency.

\subsubsection{Consensus Protocol}
The performance of blockchain is also affected by the consensus protocol.
Consensus protocols are essential components of blockchain systems, ensuring secure consensus in the presence of Byzantine adversaries. Different protocols have varying communication complexities, verifier complexities, and phases. Higher communication and verifier complexities often lead to reduced system throughput, while the number of phases affects system latency. For instance, PBFT has a communication complexity of $O(n^2)$, and the HotStuff~\cite{hotstuff} protocol, with its three-phase commit and threshold signatures, achieves linear $O(n)$ communication complexity and optimistic responsiveness. Under similar experimental conditions, the throughput and latency of these protocols also depend on whether they employ threshold signature algorithms or standard signature algorithms.

\subsubsection{Batch Size}
Batch size~\cite{abraham2020sync} influences the efficiency of block packaging and the communication overhead in the initial round, thus impacting overall system performance. In a blockchain system based on a master-slave model, transactions are packaged into blocks and then broadcast by the leader. Upon receiving the block, other nodes validate it and proceed to vote. From the first vote onwards, all subsequent phase messages contain only the block's hash rather than the entire block, improving message propagation speed and reducing storage overhead. Therefore, when node processing capacity is not exceeded, including more transactions in each block can increase system throughput, albeit at the cost of higher latency.

\subsection{Other System Fundamentals}

\subsubsection{Consensus Protocols Overview}
In our system, we have implemented three BFT consensus algorithms.

PBFT is the first BFT algorithm suitable for practical use, reducing the complexity of Byzantine fault tolerance from exponential to polynomial. PBFT is a leader-based protocol where every consensus is reached within a configuration that is called view. Each view is coordinated by a stable leader. Every node follows a ``Lock-Commit'' two-phase commit mode, handling transactions in a pessimistic manner.

HotStuff~\cite{hotstuff} is an advanced BFT protocol operating under a partially synchronous model. Adding one additional communication round to the traditional two-round protocol, HotStuff achieves linear view changes and optimistic responsiveness, allowing it to reach consensus quickly based on actual network delays rather than maximum network delay. This feature is crucial for timely transaction submissions in practical applications. HotStuff also has a variation called Chained-HotStuff., which employs a pipelined approach to achieve higher throughput. However, each view in Chained-HotStuff includes three communication rounds and packaging delays, which can increase the latency of individual transactions.

HotStuff-2~\cite{hotstuff2,zhao2024hotstuff} optimizes the HotStuff protocol by introducing an accumulated commit block mechanism, effectively reducing system latency. Additionally, it revisits the concept of optimistic responsiveness. If the leader does not have the highest lock from the previous view, it will wait for a period, which is necessary if the leader is malicious or before reaching global stability.

\subsubsection{Local Network Simulation}
We have constructed a quickly deployable local cluster architecture through local network simulation using buffered channels for communication. The modular design offers high scalability and flexibility, allowing for easy and rapid migration to realistic interconnected network environments for testing. Another advantage of local network simulation is the ability to control network latency and bandwidth freely, enabling the simulation of various network conditions ranging from low speed to high speed and from low bandwidth to high bandwidth. This flexibility ensures stability and performance under different conditions, guaranteeing the system's robustness in diverse real-world scenarios. Finally, local network simulation significantly reduces costs compared to testing in realistic network environments.
\section{Conclusion}
This paper presents a flexible private blockchain system based on the DCS trilemma—\textbf{\textit{DCS Chain}}. The core of this system lies in the definition and quantification of the DCS indicators, allowing for a multidimensional description of system performance. Through a dynamic adjustment mechanism, the system optimizes the three quantified performance indicators of the DCS trilemma, achieving theoretically optimal blockchain performance. Additionally, our system offers a complete and flexible set of core private blockchain components, making it suitable for common application scenarios. In summary, this system, grounded in the DCS trilemma theory, provides a novel solution for enhancing private blockchain performance and offers new insights for the application and development of blockchain technology.

\bibliographystyle{unsrt}
\bibliography{sample-base}

\end{document}